\documentclass[12pt]{iopart}
\usepackage{iopams}
\usepackage{amssymb}
\usepackage{graphicx}
\usepackage{epsfig}

\begin{document}

%
%
\title[Tree representation of spin-glass landscapes]{Shapes of
  tree representations of spin-glass landscapes}
\author{Wim Hordijk$^{a}$, Jos\'e F. Fontanari$^{a}$\footnote[3]{To
           whom correspondence should be addressed.}, and
        Peter F. Stadler$^{b,c,d}$}

\address{$^a$Instituto de F{\'\i}sica de S{\~a}o Carlos,
            Universidade de S{\~a}o Paulo,
            Caixa Postal 369, 13560-970 S\~ao Carlos SP, Brazil}

\address{$^b$Bioinformatik, Institut f{\"u}r Informatik,
            Universit{\"a}t Leipzig, Kreuzstra{\ss}e 7b,
	    D-04103 Leipzig, Germany}

\address{$^c$Institut f{\"u}r Theoretische Chemie und Molekulare
            Strukturbiologie, Universit{\"a}t Wien,
            W{\"a}hringerstra{\ss}e 17, A-1090 Wien, Austria}

\address{$^d$The Santa Fe Institute, 
            1399 Hyde Park Road, Santa Fe, NM 87501, USA}

\begin{abstract}

Much of the information about the multi-valley structure of disordered spin
systems can be convened in a simple tree structure -- a barrier tree -- the
leaves and internal nodes of which represent, respectively, the local
minima and the lowest energy saddles connecting those minima. Here we apply
several statistics used in the study of phylogenetic trees to barrier trees
that result from the energy landscapes of $p$-spin models.  These
statistics give information about the shape of these barrier trees, in
particular about balance and symmetry. We then ask if they can be used to
classify different types of landscapes, compare them with results obtained
from random trees, and investigate the structure of subtrees of the barrier
trees. We conclude that at least one of the used statistics is capable of
distinguishing different types of landscapes, that the barrier trees from
$p$-spin energy landscapes are quite different from random trees, and that
subtrees of barrier trees do not reflect the overall tree structure, but
their structure is correlated with their ``depth'' in the tree.
\end{abstract}

\maketitle

%
%
\section{Introduction} \label{sec:Introduction}

The notion of energy (fitness) landscapes has played a crucial role in the
development of many areas of physics and biology such as disordered
systems, neural networks, combinatorial optimization problems
\cite{Mezard:87,Fischer:91}, RNA folding \cite{Schuster:94}, and
evolutionary change \cite{Kauffman:87,Kauffman:93}, to mention only a few.
In particular, considerable effort has been devoted to the study of the
interplay between the geometry of the landscape and the nature of the
relaxation dynamics, searching heuristic or evolutionary process unfolding
on the landscape.  However, the inherent high-dimensionality of these
landscapes poses a serious hindrance to the characterization of their
topology. In fact, most of the studies have focused on the statistical
characterization of a few local properties of the landscape by looking at,
e.g., the auto-correlation function of unbiased walks over the
configuration space \cite{Weinberger:90} or the energy distribution of
local minima \cite{Bray:80,Gross:84,Oliveira:97}, while a satisfactory
description of a landscape should address also the (relative) energy
differences of the local minima, the height of the barriers between these
minima, as well as the distribution of saddle-points
\cite{Nemoto:88,Vertechi:89}.

\begin{figure}
  \begin{center}
  \includegraphics[width=4in,height=3in]{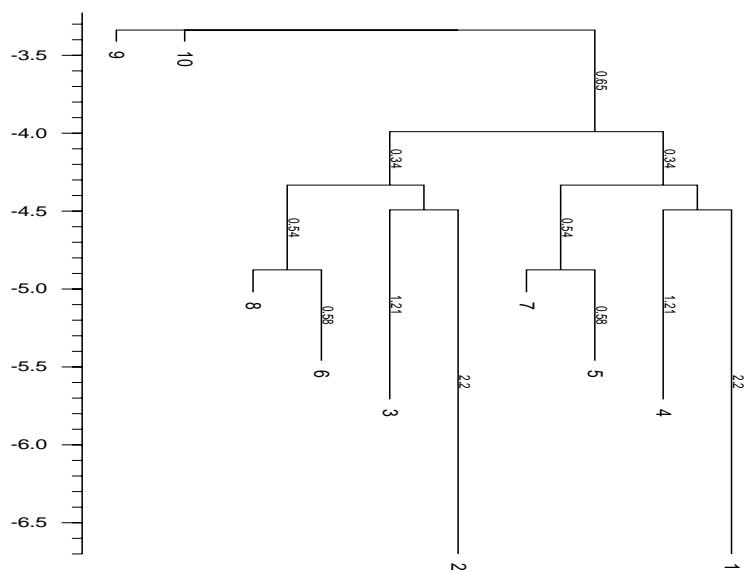}
  \end{center}
  \caption{Barrier tree for $N=10$ and $p=2$. The local minima are labeled
           1 to 10, and the height of the barriers is shown along some of the
           branches. The actual energy values can be read from the scale on
           the left.}
  \label{fig:btree}
\end{figure}

The idea of condensing all the landscape information into a tree structure,
termed {\it barrier tree}, was introduced in the context of RNA and protein
folding \cite{Becker:97,Wales:98,Garstecki:99,Flamm:00a}, and spin-glass
models \cite{Nemoto:88,Klotz:94a,Ferreira:00a,Fontanari:02a}. The advantage
of barrier trees, whose leaves represent the local minima and the internal
nodes the lowest-energy saddles connecting those minima, is that they are
both visually appealing (much information can be obtained from just looking
at them, see figure \ref{fig:btree}), as well as mathematically
well-defined, lending themselves to rigorous analysis
\cite{Flamm:02a}. However, a general quantitative measure to characterize
unambiguously different kinds of barrier trees (and hence energy
landscapes) remains to be obtained. For instance, the size-frequency
distribution of low energy saddles $ \psi (w) \sim w^{-D}$, where $w=w(s)$
is the fraction of minima that can be connected through saddle $s$, does
not provide a good measure because $D \sim 2$ regardless of whether the
barrier tree results from a spin-glass landscape or whether it is generated
randomly \cite{Fontanari:02a}.

The situation seems to be similar in the analysis of phylogenetic trees,
where it is also believed that the shape of the tree contains valuable
clues about the evolutionary process \cite{Mooers:97,Felsenstein:02}, but
so far no \emph{single} satisfactory measure of tree-shape has been
constructed \cite{Colless:82,Shao:90,Kirkpatrick:93,Fusco:95,Purvis:02a}.
In this contribution we apply five measures of tree shape that were
originally used to study phylogenetic trees (see, e.g.,
\cite{Kirkpatrick:93}) to barrier trees resulting from the Ising $p$-spin
model. These measures provide statistical information about the shape of
the barrier tree, mainly its symmetry or balance, but ignore the lengths of
the branches, i.e., the height of the barriers between minima. While the
extreme statistics of these heights provide useful information on the
performance of local search algorithms such as simulated annealing
\cite{Ferreira:00a}, a measure based solely on the shape of the barrier
tree seems more adequate to classification purposes since the shape is
probably insensitive to variations in minor details of the underlying
energy landscape.  Tree shapes are important in the biological context also
because many methods of phylogeny estimation, including parsimony, do not
produce branch lengths.  We find that all five measures can be used to tell
random from spin-glass trees, but only one measure can distinguish between
different spin-glass trees.

A few cautionary remarks regarding the relevance of the concept of energy
landscape to the understanding of the thermodynamic properties of spin
glasses are in order. The characterization of an energy landscape is
usually based on a few key elements -- local minima, energy barriers and
saddles -- which, in turn, are rigorously defined in terms of the spin
configuration space, the energy assignment to each spin configuration, and
the neighborhood relation in the configuration space (see, e.g.,
Sec. \ref{sec:EnergyLandscapes}).  However, there seems to be no simple
relation between those elements and the thermodynamic phases (equilibrium
states) of the spin-glass model, since these phases are clearly independent
of the neighborhood relation between spin configurations or the relaxation
dynamics.  To illustrate this point, we note that though the landscape of
the Ising spin glass with short-ranged interactions in three dimensions is
as complex as its counterpart of infinite-range interactions
\cite{Klotz:94a}, its thermodynamics can be successfully studied under the
assumption of a trivial ergodicity breaking at low temperature, in which
there is a single thermodynamic phase. (We refer the reader to
Ref. \cite{Fischer:91} for a lucid discussion of these controversial
issues.) Nonetheless, the study of the organization of the metastable
states (local minima) of spin glass models has a long tradition in the
physics literature of disordered systems, beginning with the work of Bray
and Moore more than two decades ago \cite{Bray:80}. It is from this
perspective that the statistical studies of energy landscapes in general,
and the present work in particular, should be considered.

In the next section, a brief overview of the Ising $p$-spin model and
energy landscapes is given.  Section \ref{sec:BarrierTrees} then reviews
the notion of barrier trees. In section \ref{sec:PhylogeneticTrees},
phylogenetic trees and several tree shape statistics used to study them are
discussed. In particular, we point out that the subtree that connects two
leaves in a barrier tree corresponds to the evolutionary path of minimum
fitness cost, in contrast with the traditional phylogenetic trees for which
the subtrees are determined by the similarity between the leaves,
regardless of the barrier height between them.  The results of applying
these measures to barrier trees of $p$-spin energy landscapes are presented
in section \ref{sec:Results}. The main conclusions are then summarized in
the final section of the paper.

%
%
\section{Energy landscapes of $p$-spin models}
\label{sec:EnergyLandscapes}

Consider a system of $N$ Ising spins $s = (s_1, \ldots, s_N)$ where $s \in
\{-1,+1\}$, with the following energy function
\begin{equation}
H_p(s) = - \sum_{1 \leq i_1 < i_2 \ldots < i_p \leq N}
  J_{i_1 i_2 \ldots i_p} s_{i_1} s_{i_2}  \ldots s_{i_p} .
\end{equation}
Here, $1 \leq p \leq N$ and the $J_{i_1 i_2 \ldots i_p}$ are i.i.d.\ random
variables from a Gaussian distribution with mean 0 and variance $p
~!/(2N^{p-1})$. Thus, each of the $2^N$ possible spin configurations is
assigned an energy value that is completely specified by the (magnetic)
interactions within all possible subsets of $p$ spins. This spin glass is
known as the $p$-spin model \cite{Gross:84,Derrida:81}. In the regime of
large $p$ and $N$ with $N \gg p$ the energies of any two distinct spin
configurations become independent random variables, distributed by a
Gaussian of mean $0$ and variance $N$, so that the random energy model
(REM) is recovered in this limit \cite{Gross:84,Derrida:81}. Here the
scaling of the variance with $N$ guarantees the extensivity of the
free-energy.  Another important limiting case is $p=2$, which corresponds
to the Sherrington-Kirkpatrick (SK) model \cite{Sherrington:75}.
 
We call two spin configurations $s$ and $t$ \emph{neighbors} if they differ
in only one of the $N$ spins, i.e., when $\sum_{i=1}^N
\left|s_i-t_i\right|=2$. In other words, neighboring configurations can be
turned into each other by a single spin flip. When $p$ is small,
neighboring spin configurations will have highly correlated energy values,
since the one spin $s_i$ that is different between the two configurations
only influences a small number of the possible subsets of $p$ spins. When
$p$ increases, the energy values of neighboring spin configurations will
become less correlated, becoming completely uncorrelated in the limit $p
\to \infty$.

The energy landscape of a $p$-spin model consists of the configuration
space $V$ of the $2^N$ possible spin configurations, with the single spin
flip neighborhood relation imposed on it, and where the energy value
$H_p(s)$ of each spin configuration $s$ is considered to be its ``height''.
This gives rise to the intuitive image of a more or less mountainous
landscape with peaks, valleys, and saddle points. A ``walk'' on this
landscape consists of moving from one neighboring spin configuration to
another, climbing up a peak or going down a valley, or perhaps just moving
around randomly. A local search algorithm such as simulated annealing can
be seen as performing such a walk, in search of the lowest valley. In
addition, $p$-spin landscapes have been used to model evolutionary
processes (see, e.g., \cite{Anderson:83,Rokhsar:86,Amitrano:89}) since they
form a class of tunably rugged landscapes similar to Kauffman's Nk-model
\cite{Kauffman:87}.  In this context, evolution is described as an
``adaptive'' walk on the energy landscape.

%
%
\section{Barrier trees} \label{sec:BarrierTrees}

A {\em local minimum} in a $p$-spin energy landscape is simply a spin
configuration $s$ that has a lower energy than all of its neighbors. A
\emph{path} $\vec{p}_{st}$ between two configurations $s$ and $t$ is a
sequence of neighboring configurations, starting at $s$ and ending at
$t$. In other words, it represents a series of single spin flips that
transforms configuration $s$ into $t$. Note that there exist multiple paths
between any pair of configurations $s$ and $t$. A \emph{saddle point}
between two local minima $s$ and $t$ is then defined as the minimum from
the set of maximum energy values along each possible path $\vec{p}_{st}$
between $s$ and $t$; see e.g.\ \cite{Nemoto:88,Vertechi:89}. So, the energy
value $E[s,t]$ of this saddle point is
\begin{equation}
E[s,t] = \min_{\vec{p}_{st}} \left\{ \max_{z \in \vec{p}_{st}} \left\{
   H_p(z) \right \} \right\} .
\end{equation}
The {\em barrier} $B(s)$ of a local minimum $s$ is defined as the height of
the lowest saddle point that connects $s$ with a local minimum $t$ of lower
energy,
\begin{equation}
B(s) = \min_t \left \{ E[s,t] - H_p(s) | H_p(t) < H_p(s) \right \} .
\end{equation}
The information about the energy values of a landscape's local minima and
the barriers that connect these local minima can be represented by a {\em
barrier tree}. In such a tree, the leaves of the tree represent the local
minima, and the internal nodes represent the saddles, with the barrier
sizes given by the length of the branches connecting the local minima to
their corresponding saddles.  Figure \ref{fig:btree} shows an example of a
barrier tree for an $N=10$ and $p=2$ $p$-spin landscape. There are 10 local
minima in this landscape (labeled 1 to 10 in the tree), with 9 saddle
points (the internal nodes). The length of each branch in the tree
indicates the height of the corresponding barrier (this value is shown
along the branch).

The algorithm for constructing these barrier trees is presented in
\cite{Flamm:00a,Flamm:02a}. It is implemented in the \texttt{barriers}
program\footnote{The source code is available at
\texttt{http://www.tbi.univie.ac.at/$\sim$ivo/RNA/Barriers/}.}, which
constructs the tree from a sorted list of energy values of all spin
configurations in the landscape. The program \texttt{barriers} is used here
to generate barrier trees of $p$-spin landscapes, see also
\cite{Ferreira:00a,Fontanari:02a} for applications to spin glass problems.

%
%
\section{Phylogenetic trees}   \label{sec:PhylogeneticTrees}

Phylogenetic trees are often used to study the historical relations within
or between groups of biological species; see e.g.\
\cite{Felsenstein:02}. The currently existing species (or subspecies) form
the leaves of the tree, and two related species are linked through their
last common ancestor, which forms an internal node in the tree. The length
of the branch between a species and its ancestor indicates how long ago the
speciation event occurred that led to the current species. ``Dead-ends'' in
this tree represent extinction events. So, the shape of a phylogenetic tree
contains information about patterns of speciation and extinction (and
possibly about the rates of these occurrences), and thus tells us something
about the evolution of different species.

\begin{figure}
  \begin{center}
  \includegraphics[width=4in,height=3in]{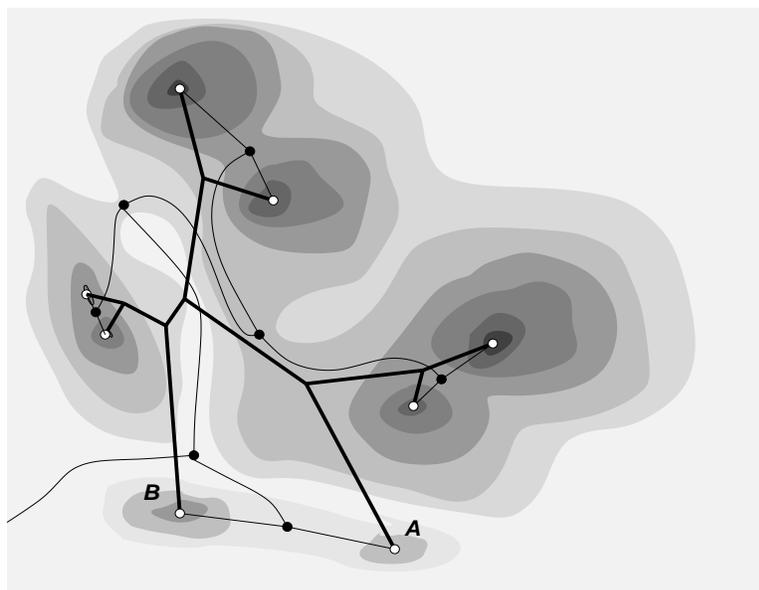}
  \end{center}
  \caption{The tree obtained from the minimum fitness paths, shown by thin
lines with the saddle points indicated by black dots and the leaves by
white dots, does not necessarily coincide with the trees obtained from
clustering methods that are based on sequence similarity, shown here with
thick lines. The darker regions indicate higher fitness configurations.}
  \label{fig:arbol}
\end{figure}

The traditional approaches to reconstruct or infer a phylogenetic tree
given the extant species (i.e., the leaves) are based on sequence
similarity, i.e., configurational overlap only, which can be justified by
the (usually implicit) assumptions of a flat fitness landscape and a
diffusive behavior in sequence space \cite{Felsenstein:02}.  However, it
seems intuitive that, regardless of the similarity between two sequences,
if the fitness costs of all possible evolutionary paths connecting them are
high, they must be put far apart in the phylogenetic tree. In the barrier
tree approach both similarity and fitness cost are taken into account to
yield the evolutionary path of {\it minimum fitness cost} connecting two
species.  Thus it may be viewed as a generalization of the maximum
parsimony principle to rugged fitness landscapes. We recall that maximum
parsimony chooses the tree (or trees) that require the fewest evolutionary
changes (spin flips, in the present context).  In figure \ref{fig:arbol} we
give an example in which the distance-based tree and barrier tree have
different topologies.
 
Various methods have been proposed to analyze phylogenetic trees. Here, we
consider five statistics that were used in \cite{Kirkpatrick:93} to measure
tree symmetry and balance. The trees are assumed to be binary trees with
$n$ leaves (or species) and thus $n-1$ internal nodes, with the root being
the last common ancestor of all $n$ species. Let $d(i,j)$ be the
graph-theoretical distance between two nodes of the tree, i.e., the number
of edges along the path that connects them. Furthermore, we denote the root
of the tree by $\varnothing$. The \emph{height} of a leaf $l$ is
$h_l=d(\varnothing,l)$. Equivalently, $h_l$ is the number of internal nodes
between leaf $l$ and the root $\varnothing$ (inclusive). For each interior
node $i$ we have two subtrees with $r_i$ and $s_i$ leaves, respectively. We
assume $r_i\geq s_i$. The \emph{subtree-height} of an interior node $i$ is
$m_i=\max_{l\in T_i} d(i,l)$ where the maximum is taken over all leaves $l$
in the subtree $T_i$ below $i$, i.e., the subtree of which $i$ is the root.

With this notation we may define the following five characteristic values
for the shape of a binary rooted tree:
\begin{enumerate}
\item $H = \frac{1}{n} \sum_{l=1}^n h_l$ is the \emph{average height} of a
      leaf in the tree\footnote{In \cite{Kirkpatrick:93} this quantity is
      denoted by $\bar N$.}.
\item $\sigma_H = \sqrt{\frac{1}{n} \sum_{l=1}^n \left(h_l-H\right)^2}$,
      is the standard deviation of the leaf height.
\item $C = \frac{2}{n(n-3)+2} \sum_{i=1}^{n-1} \left(r_i-s_i\right)$ is a
      measure for the \emph{imbalance} of the tree. Up to normalization it
      is the same as Colless's imbalance measure \cite{Colless:82}.  A
      closely related measure of this type, which we will not use here,
      essentially amounts to averaging $(r_i-s_i)/(r_i+s_i)$ instead,
      see e.g.\ \cite{Fusco:95,Purvis:02a}.
\item $B_1 = \sum_{i\ne\varnothing} 1/m_i$ is the average inverse
      subtree height, where the sum is taken over all $n-2$ internal nodes
      $i$ excluding the root $\varnothing$.
\item $B_2 = \sum_{l=1}^n 2^{-h_l}h_l$ is an
      alternatively weighted average leaf-height.
\end{enumerate}
In \cite{Kirkpatrick:93} the variance $\sigma_H^2$ was considered, but here
we will use the standard deviation. Both $H$ and $\sigma_H$ have larger
values for more asymmetric trees. In \cite{Kirkpatrick:93} it is shown that
the expected value of $\langle H\rangle = 2\sum_{k=2}^n 1/k$ for random
trees with $n$ leaves. Values of $H$ larger than this indicate trees more
asymmetric than a random tree.  For a completely symmetric tree,
$\sigma_H=0$, while it has a maximum value for a completely asymmetric
tree.

The imbalance measure $C$ examines the internal nodes of a tree. It
``weighs'' the subtrees branching out from each internal node by
counting and comparing the number of leaves in each subtree. These
weight differences are then averaged and normalized over all internal
nodes of the tree. The value of $C$ increases from 0 for a completely
symmetric tree to 1 for a completely asymmetric tree.  The quantity
$B_1$ looks at the longest possible path $m_i$ from each internal node
$i$ to any of the leaves in its subtree.  The statistic $B_2$ is based
on an index of information content. For highly asymmetric trees $B_2$
will quickly converge to a value of 2. For a completely symmetric
tree, it will be equal to $\log_2(n)$, where $n$ is the number of
leaves in the tree. Both $B_1$ and $B_2$ have smaller values for
increasingly asymmetric trees.

%
%
\section{Results} \label{sec:Results}

Here, we apply the statistics presented in the previous section to barrier
trees that result from $p$-spin energy landscapes. The parameter values
used are $N=10$, $12$, $15$, $18$, $20$, and $22$, and $p=2$, $3$, and
$\infty$ (REM). For each combination of $N$ and $p$, $100$ independent
landscapes were generated randomly, and the barrier trees for each of these
landscapes were constructed. The tree statistics reported here are the
averages of the 100 trees for each parameter combination.

\begin{figure}
  \begin{center}
    \begin{tabular}{ccc}
      \includegraphics[width=0.3\textwidth,clip=]{N.eps} &
      \includegraphics[width=0.3\textwidth,clip=]{C.eps} &
      \includegraphics[width=0.3\textwidth]{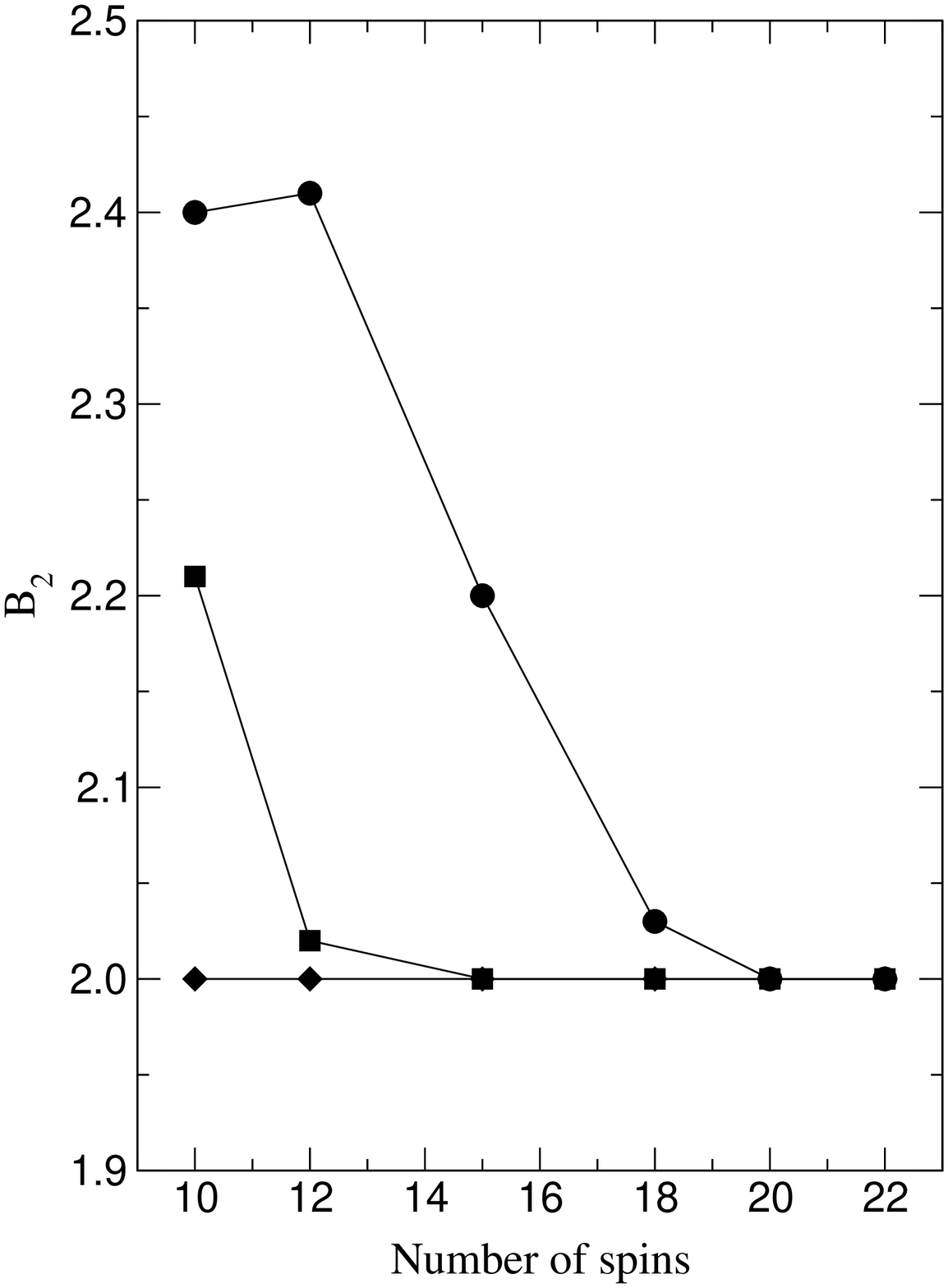} \\
		      {\small(a)} & {\small(b)} & {\small(c)} 
    \end{tabular}
  \end{center}
  \caption{Size dependence of tree statistics as a function of the number
    of spins $N$ for three classes of landscapes:
    $p$-spin models with $p=2$ ($\bullet$) and $p=3$ ($\blacksquare$), and
    the REM ($\blacklozenge$). Each data point is an
    average over 100 independent landscapes. The solid line is an
    exponential fit to the data for the average leaf height $H$ in (a).
    The solid lines in (b) are fits to $C = a + b \ln N$. We do not have
    a justification for this particular functional form.
    The values of $B_2$ quickly converge to the asymptotic value of $2$
    in (c), where the solid lines are simply guides to the eye.}
  \label{fig:NCB}
\end{figure}

Three of the statistics ($H$, $\sigma_H$, and $B_1$) are exponential in
$N$, the number of spins. For example, the relation between $H$ and $N$ is
$H=0.43\exp(0.20N)$ for $p=2$.  The data for $H$ (symbols) and the
exponential fits (solid lines) are shown in a semi-log plot in figure
\ref{fig:NCB}a for the different values of $p$. The data for $\sigma_H$ and
$B_1$ (not shown) are similar, but with different slopes. It turns out that
the average number of leaves, $n$, in the barrier trees is also exponential
in the number of spins $N$, with the same slope as for $H$. Analytical
values for this exponent, defined as $\lim_{N\to\infty} \ln n/N$, for the
$p$-spin models are derived e.g.\ in \cite{Gross:84} (see also
\cite{Oliveira:99a}). The numerical values from the simulations reported
here are compiled in table \ref{tab:slopes}.

The imbalance $C$ is sub-linear in $N$, as shown in figure \ref{fig:NCB}b,
and eventually converges to $1$ for large $N$ (so the logarithm fitting
cannot remain valid for all $N$). However, the value of $B_2$ very quickly
converges to $2$ with increasing $N$, as shown in Figure \ref{fig:NCB}c,
and thus it does not seem to be a very useful measure to distinguish the
trees. It should be noted at this point that the number $n$ of leaves,
i.e., local minima, depends not only on $N$ but also on the type of the
landscape. We will therefore consider tree measures as a function of $n$
rather than $N$ below. In fact, $n$ is the natural parameter in the
analysis of phylogenetic trees, obtained by simply counting the number of
leaves in the tree.

The data presented so far explicitly assumed that we already know the
values of $N$ and $p$ of the energy landscapes. Suppose we do not know
these values, but all we have are several instances (say, $100$) of barrier
trees that were derived from some $p$-spin model. Is it possible to say
something about the underlying landscape from just the barrier tree? (The
problem faced by the phylogeneticists is even harder since they have access
to a single tree of arbitrary size only \cite{Kirkpatrick:93}.)

\begin{table}[b]
  \begin{center}
    \begin{tabular}{lrrr}
      \hline\hline
      &  $p=2$ &  $p=3$ &    REM \\
      \hline
      $H$        & 0.1999 & 0.3157 & 0.6340 \\
      $\sigma_H$ & 0.2485 & 0.3381 & 0.6374 \\
      $B_1$      & 0.1070 & 0.1832 & 0.6040 \\
      $n$        & 0.1951 & 0.2945 & 0.6306 \\
      \hline\hline
    \end{tabular}
    \caption{The slopes of the exponential fits for the statistics
      $H$, $\sigma_H$, $B_1$, and $n$,
      for each value of $p$.}
    \label{tab:slopes}
  \end{center}
\end{table}

\begin{figure}
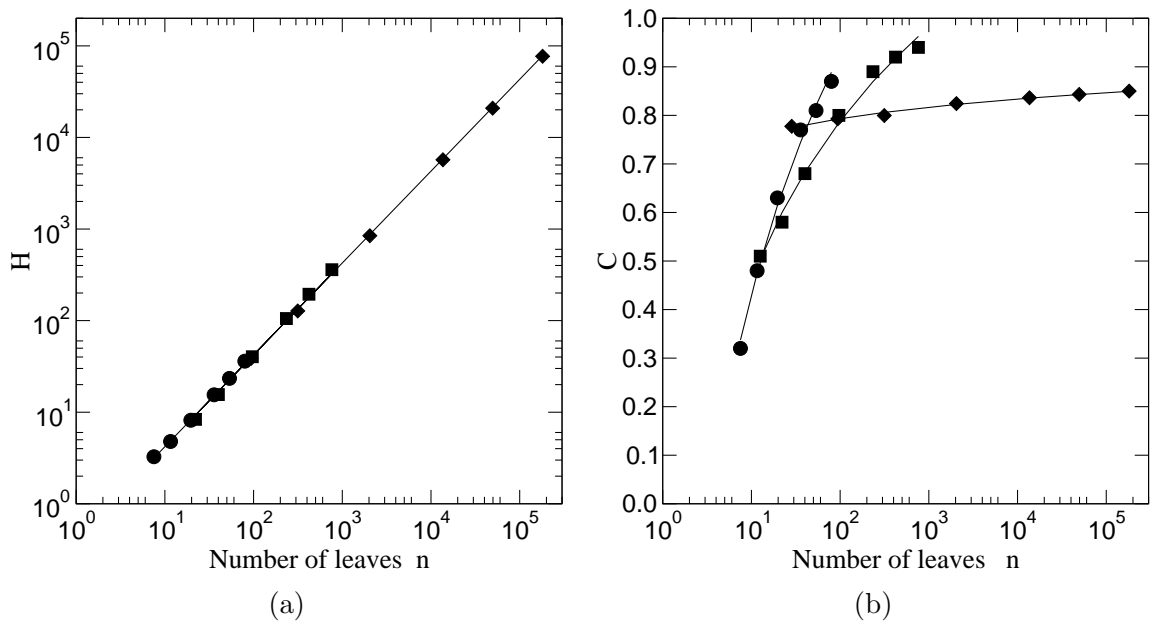

  \begin{center}
    \begin{tabular}{cc}
      \includegraphics[width=0.47\textwidth,clip=]{n.N.eps}  &
      \includegraphics[width=0.47\textwidth,clip=]{n.C.eps}  \\
		      {\small(a)} & {\small(b)}
    \end{tabular}
  \end{center}
  \caption{(a) The values of $H$ and (b) the values of $C$ plotted against
    $n$, the average number of leaves in the tree. We find that $H$ does
    not depend on the $p$ for different $p$-spin models.  On the other
    hand, the values of $C$ show a significant dependence on the details of
    the landscape. The fit shown is $C = a + b \ln \ln n$ and the
    convention is\break $\bullet$ $p=2$, $\blacksquare$ $p=3$, and
    $\blacklozenge$ REM.}
  \label{fig:n_}
\end{figure}

In figure \ref{fig:n_}a, the values of $H$ of all landscapes are lumped
together and plotted against $n$, the (average) number of leaves in the
tree, in a log-log plot. This results in a straight line, indicating a
power-law behavior. In fact, the slope in this case is 1, since the slopes
of $H$ and $n$ when plotted against $N$ are equal (see table
\ref{tab:slopes}).  The plots for $\sigma_H$ and $B_1$ (not shown) are
similar, but with slopes slightly larger and slightly smaller than 1,
respectively.

In this case, it is not possible to distinguish the trees that result from,
e.g., the $p=2$ and the $p=3$ landscapes: they all fall on the same
line. Since $B_2$ converges to a fixed value for larger trees, it is not
very useful for distinguishing different trees either. However, when the
imbalance statistic $C$ is plotted against $n$, there is a clear
distinction, as can be seen in figure \ref{fig:n_}b. In particular, if one plots
 $C$ against $\ln \ln n$ (i.e., if $n$ is on a double logarithmic scale), the data is fitted
by straight lines with different slopes for the different values of
$p$ (data not shown). So, $C$ clearly can be used as a statistic to distinguish and classify
different barrier trees, and thus their underlying landscapes.

Next we address the question: How different are barrier trees of $p$-spin
landscapes from random trees? To answer it, we generated random binary
trees with $n=10^i$ leaves for $i=1$, $2$, $3$, $4$, and $5$. For each
value of $i$, 100 random trees were generated and the same five statistics
were calculated.  The random trees were generated as follows. First, create
$n$ nodes (the leaves) and put them in a set $A$. Next, remove two random
nodes $x$ and $y$ from $A$, create a new node $z$ and make $x$ and $y$ its
two children, and put $z$ in the set $A$. Repeat this procedure until there
is only one node left in $A$, which will be the root of the tree.

\begin{figure}
  \begin{center}
    \begin{tabular}{cc}
      \includegraphics[width=0.47\textwidth,clip=]{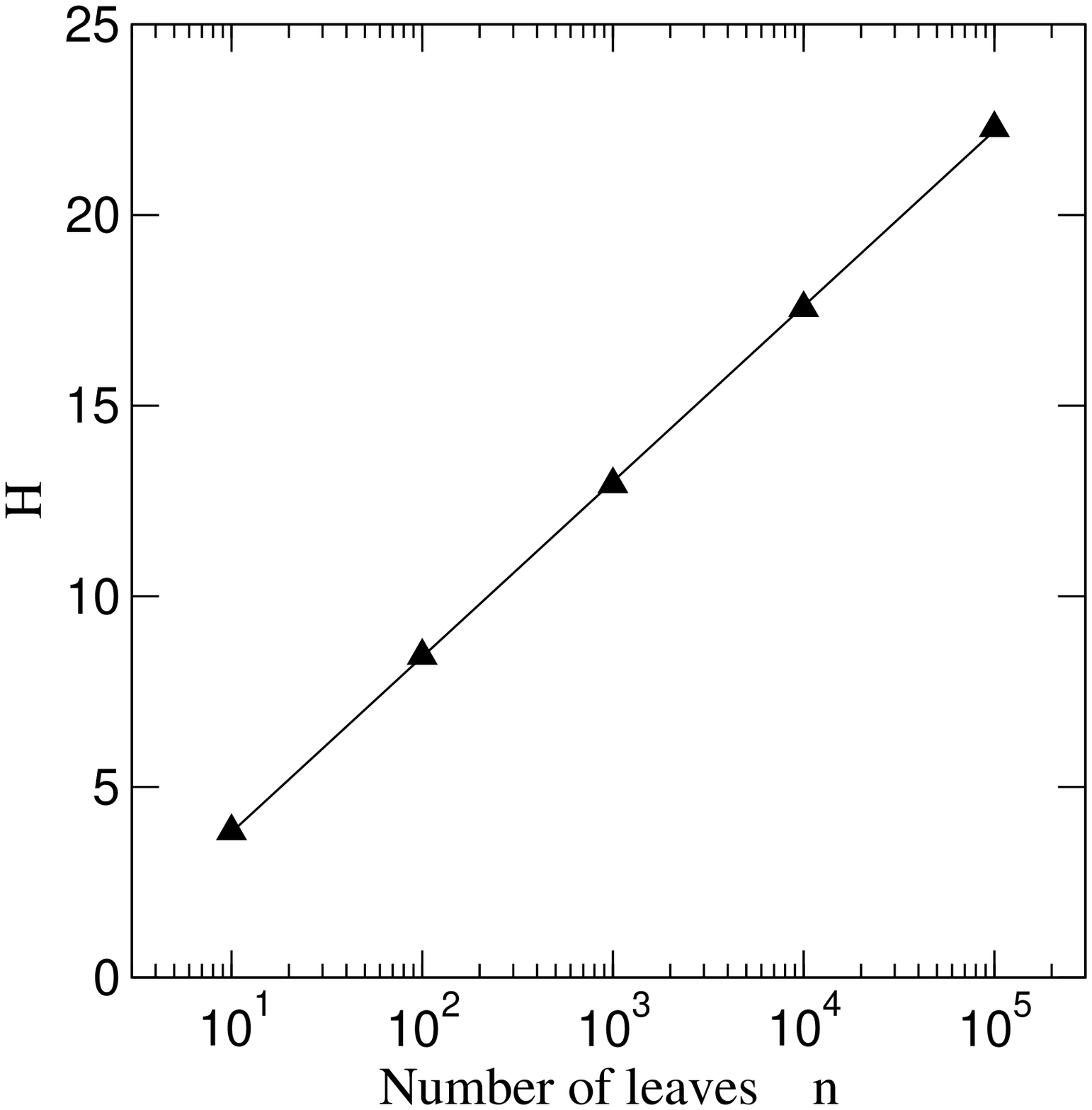} &
      \includegraphics[width=0.47\textwidth,clip=]{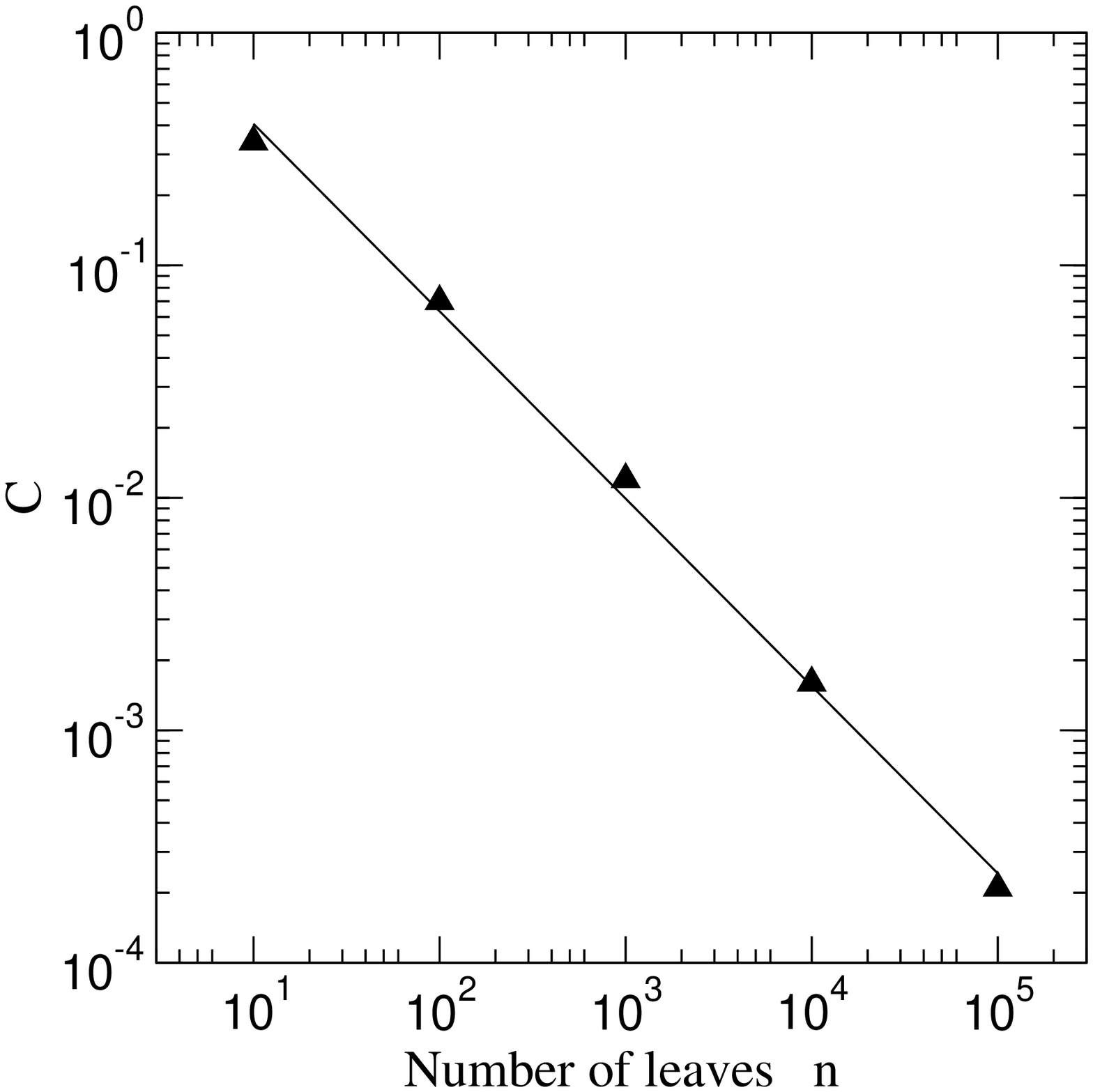} \\
      {\small(a)} & {\small(b)} 
    \end{tabular}
  \end{center}
  \caption{Average leaf-height $H$ and imbalance measure $C$ as a function of
    tree-size $n$ for random trees.}
  \label{fig:rnd_NC}
\end{figure}

Figure \ref{fig:rnd_NC}a shows the data for $H$ against $n$ (the number of
leaves) in a semi-log plot for random trees. Clearly, $H$ depends
logarithmically on $n$. The results for $\sigma_H$ and $B_2$ (not shown)
are similar, but with different slopes. Figure \ref{fig:rnd_NC}b shows the
data for $C$ against $n$ on a log-log plot. The fit to the data is a power
law with exponent $-0.81$. The results for $B_1$ (not shown) are similar
but with a (positive) exponent very close to 1.

The results for random trees are quite different from those for the
barriers trees of $p$-spin landscapes. In the $p$-spin case, $H$ and
$\sigma_H$ depend exponentially on $n$, whereas for random trees this
dependence is logarithmic. For $C$, the reverse is true. Moreover, for
random trees, $C$ {\em decreases} with increasing $n$, while it increases
for $p$-spin barrier trees. For $B_1$, in both cases the dependence is
exponential, but with different exponents.  Finally, for $B_2$ both cases
are completely different, with a (downward) convergence to the value $B_2
=2$ for $p$-spin trees, and a logarithmic increase for random
trees. Clearly, barrier trees from $p$-spin landscapes are much more
asymmetric than random trees. A similar, though qualitative, conclusion was
reached by considering the size-distribution of minima connected through a
high-energy (i.e., closer to the root) saddle-point \cite{Fontanari:02a}.

Finally, we investigate the structure of subtrees of barrier trees. Instead
of calculating the tree statistics on the entire tree, they are calculated
on subtrees starting at some internal node of the tree. This way, it can be
determined whether the tree has a self-similar structure or not (i.e.,
whether subtrees look similar to the tree as a whole).  Three different
instances of an $N=20, p=3$ $p$-spin landscape were taken and their barrier
trees were constructed. Next, the five statistics were calculated on the
subtrees starting from each of the $n-1$ internal nodes of these trees (a
binary tree with $n$ leaves has $n-1$ internal nodes). Each internal node
in a barrier tree represents a saddle point in the underlying energy
landscape, and the internal nodes are characterized by the energy values of
their corresponding saddle points. Obviously, nodes higher up in the tree
(closer to the root) will have higher energy values than nodes lower down
(or deeper) in the tree (closer to the leaves).

\begin{figure}
  \begin{center}
    \begin{tabular}{cc}
      \includegraphics[width=0.47\textwidth,clip=]{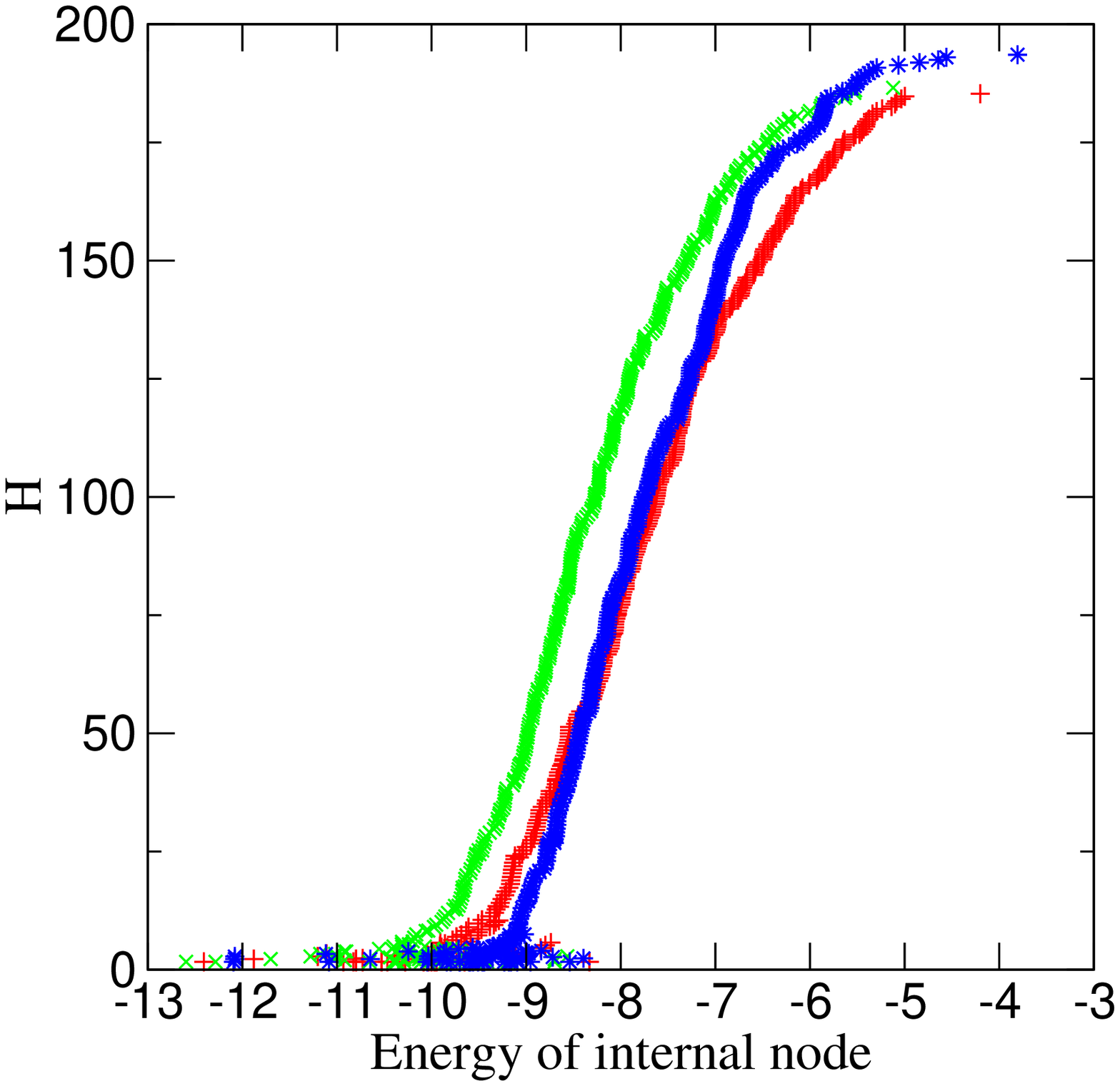} &
      \includegraphics[width=0.47\textwidth,clip=]{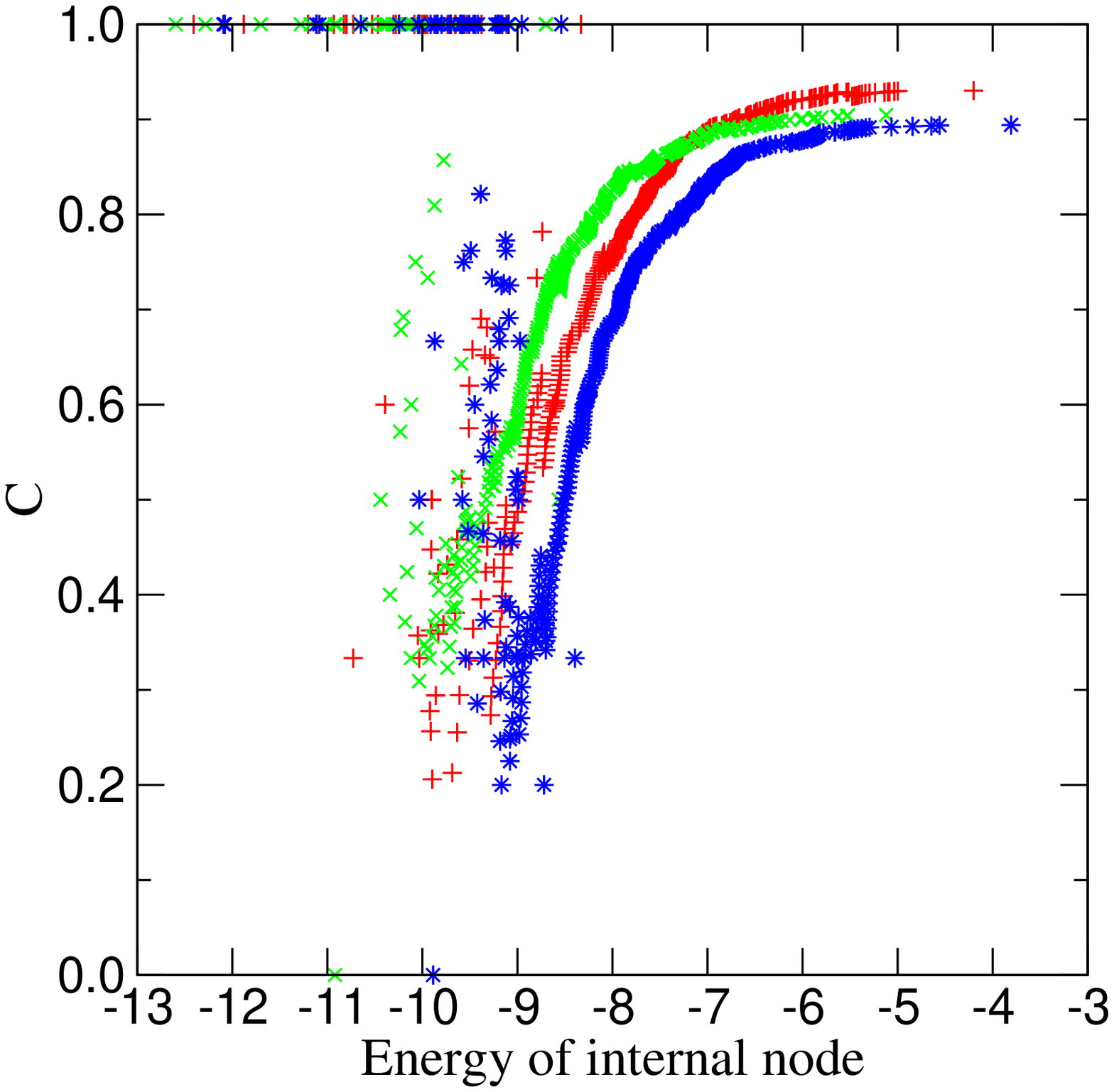} \\
      {\small(a)} & {\small(b)} 
    \end{tabular}
  \end{center}
  \caption{(a) $H$ and (b) $C$ against the energy of the
    internal node (saddle point)
    for three different landscapes for $N=20$ and $p=3$.}
 \label{fig:subtree}
\end{figure}

Figure \ref{fig:subtree}a shows the results for $H$. The statistics for the
subtrees are plotted against the energy value of the internal node that
forms the root of the subtree. The data for the three different landscapes
can be distinguished from the slightly different range in energy values,
but the overall shape is the same for all three. Clearly, the values for
$H$ depend strongly on the energy value of the root of the subtree. The
results for $\sigma_H$ and $B_1$ (not shown) are similar. Figure
\ref{fig:subtree}b shows the results for the statistic $C$, where we have
discarded subtrees with $n=2$ since $C$ is not defined for them. We note
that $C=1$ for all subtrees with $n=3$ leaves. We find that subtrees with a
high-energy root are extremely unbalanced. The same is true for the
subtrees with very few leaves that we find near the global optimum. In the
intermediate regime we find nodes with very balanced subtrees. For $B_2$,
the values vary much more widely, but with most of the points falling on or
near the $B_2=2$ line. The overall results for $p=2$ and for the REM are
very similar.

>From these plots it is clear that the structure of subtrees does not
reflect the structure of the tree as a whole. Instead, there is a rather
well-defined dependence on the structure of a subtree and its \emph{depth}
in the tree, i.e., the energy value of the internal node that forms its
root. The landscape structure around local minima and saddle points with a
relatively high energy value is therefore significantly different from the
structure around local minima and saddle points with a low energy. Since
this structure is correlated with the energy value of the local minima and
saddle points, this information could possibly be used to guide local
search algorithms.

%
%
\section{Conclusions} \label{sec:Conclusions}

One of the main outcomes of the mean-field (replica) theory of spin glasses
was the prediction that, in the low-temperature phase, the phase space is
broken into infinitely many pure states or valleys
\cite{Mezard:87,Fischer:91}. Finding universal, in the sense of model
independent, features of the distribution of valleys is one of the main
goals of the theory \cite{Derrida:97,Bouchaud:97}. Unfortunately, the
concept of valley in the replica theory is not easily related to the more
tangible concepts of local minima and saddles commonly used to characterize
complex landscapes. The reason being that the replica valleys are weighted
by their Boltzmann factors which, at least for Gaussian distributed
couplings, results in a single valley (up to an overall spin flip) at zero
temperature, thus contrasting with the exponentially large number of local
minima of the landscape. In that sense, the replica theory is of little use
in the characterization of energy landscapes of spin-glass models.  To
investigate the organization of the local minima one has then to resort
either to the annealed estimates of the correlations between local minima
in the thermodynamic limit \cite{Gross:84,Oliveira:97} or to the exact
numerical calculation of the barrier trees for relatively small system
sizes \cite{Nemoto:88,Vertechi:89,Fontanari:02a}. As pointed out before,
the latter seems a more convenient approach for the purpose of classifying
families of landscapes or spin-glass models. Furthermore, the barrier tree
also contains information about the structure of subspaces in the landscape
around local minima and saddle points, since the structure and symmetry of
the subtrees are clearly correlated with their depth in the tree (or the
energy level of the node that forms the root of the subtree). This
information can probably be used to guide local search algorithms such as
simulated annealing in their search for the lowest (or global) minimum.

In searching for efficient measures to characterize landscapes, it was
recently shown that the size-frequency distribution of the number of leaves
$w$ connected by a saddle-point $s$ is too robust to be a useful measure,
as it yields the same power-law $\psi (w) \sim w^{-2}$ for both $p$-spin
barrier trees and random trees \cite{Fontanari:02a}. In this contribution,
we improve considerably the landscape systematics by considering five
measures proposed originally to characterize the shape of phylogenetic
trees \cite{Kirkpatrick:93}. Three of the measures used, namely, $H$,
$\sigma_H$ and $B_1$ were proved independent of the underlying spin-glass
landscape when plotted against the number of leaves $n$, as evidenced by
the ``data collapse'' illustrated in figure \ref{fig:n_}a. Only one of the
measures, the imbalance $C$, can be used to differentiate between, e.g.,
$p=2$-type and $p=3$-type energy landscapes.  It is remarkable, however,
that all five measures yield completely different results for random trees,
owing to a different scaling with $n$. In particular, barrier trees
generated from energy landscapes of $p$-spin models appear to be much more
asymmetric than random trees, and the asymmetry increases with increasing
$N$ (the number of spins) or $n$ (the number of leaves). It remains a
challenge to find (if it exists) a disordered spin system whose associated
barrier trees exhibit balance and symmetry properties akin to those of the
random trees or, perhaps an easier task, that violate the scaling law shown
in figure \ref{fig:n_}a.

%
%
\section*{Acknowledgments}

Thanks to Christoph Flamm at the University of Vienna, Austria, for his
help with the \texttt{barriers} program. This research was supported by
Funda\c{c}\~ao de Amparo \`a Pesquisa do Estado de S\~ao Paulo (FAPESP),
project 99/09644-9. The work of J.F.F. is supported in part by CNPq and WH
is supported by FAPESP.

%
%


\section*{References}

\begin{thebibliography}{99}

\bibitem{Mezard:87} M\'ezard M, Parisi G and Virasoro M A 1987 {\it Spin
Glass Theory and Beyond} (Singapore: World Scientific)

\bibitem{Fischer:91} Fischer K H and Hertz J A 1991 {\it Spin Glasses}
(Cambridge: Cambridge University Press)

\bibitem{Schuster:94} Schuster P and Stadler P F 1994 {\it Comput. Chem.}
{\bf 18} 295

\bibitem{Kauffman:87} Kauffman S A and Levin S 1987 {\it J. Theor. Biol.}
{\bf 128} 11

\bibitem{Kauffman:93} Kauffman S A 1993 {\it The Origins of Order} (Oxford:
Oxford University Press)

\bibitem{Weinberger:90} Weinberger E D 1990 {\it Biol. Cybern.} {\bf 63}
325

\bibitem{Bray:80} Bray A J and Moore M A 1980 {\it J. Phys. C} {\bf 13} L469

\bibitem{Gross:84} Gross D J and M\'ezard M 1984 {\it Nucl. Phys. B} {\bf
240} 431

\bibitem{Oliveira:97} de Oliveira V M and Fontanari J F 1997 {\it
J. Phys. A: Math. Gen.} {\bf 30} 8445

\bibitem {Nemoto:88} Nemoto K 1988 {\it J. Phys. A: Math. Gen.} {\bf 21}
L287

\bibitem{Vertechi:89} Vertechi A M and Virasoro M A 1989 {\it
J. Phys. France} {\bf 50} 2325

\bibitem{Becker:97} Becker O M and Karplus M 1997 {\it J. Chem. Phys.}
{\bf 106} 1495

\bibitem{Wales:98} Wales D J, Miller M A and Walsh T R 1998 {\it Nature}
{\bf 394} 758

\bibitem{Garstecki:99} Garstecki P, Hoang T X and Cieplak M 1999 {\it
Phys. Rev. E} {\bf 60} 3219
  
\bibitem{Flamm:00a} Flamm C, Fontana W, Hofacker I L and Schuster P 2000
{\it RNA} {\bf 6} 325

\bibitem{Klotz:94a} Klotz T and Kobe S 1994 {\it J. Phys. A: Math. Gen.}
{\bf 27} L95

\bibitem{Ferreira:00a} Ferreira F F, Fontanari J F and Stadler P F 2000
{\it J. Phys. A: Math. Gen.} {\bf 33} 8635

\bibitem{Fontanari:02a} Fontanari J F and Stadler P F 2002 {\it J. Phys. A:
Math. Gen.}  {\bf 35} 1509

\bibitem{Flamm:02a} Flamm C, Hofacker I L, Stadler P F and Wolfinger M T
2002 {\it Z.  Phys. Chem.} {\bf 216} 155

\bibitem{Mooers:97} Mooers A O and Heard S B 1997 {\it Quart. Rev. Biol.}
{\bf 72} 31

\bibitem{Felsenstein:02} Felsenstein J 2002 {\it Inferring Phylogenies}
(Sunderland: Sinauer Associates)

\bibitem{Colless:82} Colless D H 1982 {\it Syst. Zool.} {\bf 31} 100
 
\bibitem{Shao:90} Shao K-T and Sokal R R 1990 {\it Syst. Zool.} {\bf 39}
266

\bibitem{Kirkpatrick:93} Kirkpatrick M and Slatkin M 1993 {\it Evolution}
{\bf 47} 1171

\bibitem{Fusco:95} Fusco G and Cronk Q C B 1995 {\it J. Theor. Biol.} {\bf
175} 235

\bibitem{Purvis:02a} Purvis A, Katzourakis A and Agapow P-M 2002 {\it
J. Theor. Biol.} {\bf 214} 99

\bibitem{Derrida:81} Derrida B 1981 {\it Phys. Rev. B} {\bf 24} 2613

\bibitem{Sherrington:75} Sherrington D and Kirkpatrick S 1975 {\it
Phys. Rev. Lett.} {\bf 35} 1792

\bibitem{Anderson:83} Anderson P W 1983 {\it Proc. Natl. Acad. Sci. USA}
{\bf 80} 3386

\bibitem{Rokhsar:86} Rokhsar D S,  Anderson P W and Stein D L 1986 
{\it J. Mol Evol.} {\bf 23} 119

\bibitem{Amitrano:89} Amitrano C, Peliti L and  Saber M 1989
{\it J. Mol. Evol.} {\bf 29} 513 

\bibitem{Oliveira:99a} de Oliveira V M, Fontanari J F and Stadler P F 1999
{\it J. Phys. A: Math. Gen.} {\bf 32} 8793

\bibitem{Derrida:97} Derrida B 1997 {\it Physica D} {\bf 107} 186

\bibitem{Bouchaud:97} Bouchaud J-P and M\'ezard M 1997
{\it J. Phys. A: Math. Gen.} {\bf 30} 7997

 
\end{thebibliography}
\end{document}